\documentclass[aps,prl,groupedaddress]{revtex4-1}  
\usepackage{graphicx}  
\usepackage{bm}        
\usepackage{amssymb,amsmath}   
\usepackage{verbatim}
\usepackage[T1]{fontenc}
\usepackage{amsmath,amssymb,amsfonts}
\usepackage{cleveref}               

\usepackage{ifpdf}
\usepackage{color}

\usepackage{amsmath}
\usepackage{graphics}
\usepackage{mathtools}
\usepackage[usenames,dvipsnames]{xcolor}
\usepackage{epsfig}
\usepackage{epstopdf}
\usepackage{dcolumn}
\usepackage[reset]{geometry} 
\makeatletter
\Gm@restore@org
\makeatother
\usepackage{tikz}
\usetikzlibrary{shapes.geometric, arrows}
\usepackage{upgreek}
\usepackage{setspace}
\usepackage{enumitem}
\usepackage{array,multirow,bigdelim}

\def\be{\begin{equation}}
\def\ee{\end{equation}}
\def\ba{\begin{eqnarray}}
\def\ea{\end{eqnarray}}

\def\dd{{\dot\delta}}

\def\CP1{\mathbb{CP}^1}
\def\SL2C{\mathrm{SL}(2,\mathbb{C})}

\def\Z2{\mathbb{Z}_2}

\def\su2{{SU(2)}}

\def\[{\left[}
\def\]{\right]}

\def\({\left(}
\def\){\right)}
\def\[{\left[}
\def\]{\right]}

\def\i2{\frac{i}{2}}

\def\2F1{\,_2{\rm F}_1}

\newcommand{\beq}{\begin{equation}}
\newcommand{\eeq}{\end{equation}}
\newcommand{\beqq}{\begin{equation*}}
\newcommand{\eeqq}{\end{equation*}}
\newcommand\beqa{\begin{eqnarray}}
\newcommand\eeqa{\end{eqnarray}}
\newcommand\beqaa{\begin{eqnarray*}}
\newcommand\eeqaa{\end{eqnarray*}}
\newcommand\bea{\begin{array}}
\newcommand\eea{\end{array}}

\def\dd{d}

\def\Riem{R}

\begin{document}

\preprint{}

\title{Scattering Angles in Kerr Metrics}
\author{Poul H. Damgaard$^1$, Jitze Hoogeveen$^1$, Andres Luna$^1$, and Justin Vines$^2$\\ }
\affiliation{\smallskip {$^1$}Niels Bohr International Academy\\ Niels Bohr Institute, University of Copenhagen\\
Blegdamsvej 17, DK-2100 Copenhagen \O, Denmark \\
{$^2$}Max Planck Institute for Gravitational Physics (Albert Einstein Institute)\\
Am Muhlenberg 1, Potsdam 14476, Germany}
\date{\today}

\begin{abstract}
Scattering angles for probes in Kerr metrics are derived for scattering in the equatorial plane of the black hole. We use a method that 
naturally resums all orders in the spin of the Kerr black hole, thus facilitating comparisons with scattering-angle computations based 
on the Post-Minkowskian expansion from scattering amplitudes or worldline calculations. We extend these results to spinning black-hole 
probes 
up to and including second order in the probe spin and any order in the Post-Minkowskian expansion, for probe spins aligned with the Kerr spin. When truncating to third Post-Minkowskian order, 
our results agree with those obtained by amplitude and worldline methods.
\end{abstract}
\pacs{04.60.-m, 04.62.+v, 04.80.Cc}
\maketitle

\section{Introduction}

The gravitational bending of light around the Sun famously provided one of the earliest observational checks on predictions from
Einstein's general theory of relativity. Since then, gravitational optics has become one of the common tools of observational astronomy, in
fact now used inversely to infer mass distributions of massive objects partly obstructing light in the direct line of sight. The light-like bending
angles of general relativity thus have a central position in modern physics. 

Recently, scattering angles of massive objects have attracted renewed attention from an entirely different direction. For gravitational-wave
predictions of black-hole mergers one needs the effective Hamiltonian that governs the dynamics of two massive bodies in general relativity.
It was suggested in Ref. \cite{Damour:2016gwp} that an improvement of traditional analytical approaches based on
Post-Newtonian expansions could come from the Post-Minkowskian expansion of the scattering regime. This suggests that modern amplitude
methods of the quantized theory to great advantage may be used to infer the effective two-body interactions of general relativity
\cite{Bjerrum-Bohr:2018xdl,Cheung:2018wkq,Cristofoli:2019neg} after properly removing all non-classical contributions at loop level
\cite{Kosower:2018adc}. In a short span
of time there has been enormous progress in this direction, with results to third Post-Minkowskian order now fully under control
\cite{Bern:2019nnu,Antonelli:2019ytb,Parra-Martinez:2020dzs,Mougiakakos:2021ckm,Herrmann:2021tct,DiVecchia:2020ymx,Damour:2020tta,Bjerrum-Bohr:2021vuf,Damgaard:2021ipf,Brandhuber:2021eyq}. Even amplitude computations to fourth Post-Minkowskian order \cite{Bern:2021dqo} and, in the probe limit, fifth Post-Minkowskian
order \cite{Bjerrum-Bohr:2021wwt} have now been considered. A parallel track based on effective field theory in the worldline formalism offers
results at similar high orders in the Post-Minkowskian expansion \cite{
Kalin:2020mvi,
Kalin:2022hph,
Kalin:2020fhe,
Liu:2021zxr,
Dlapa:2021npj,
Mogull:2020sak,
Jakobsen:2021smu,
Jakobsen:2022fcj,
Jakobsen:2022psy,Saketh:2022wap}. 
For recent reviews, see, $e.g.$, refs. \cite{Bjerrum-Bohr:2022blt,Kosower:2022yvp,Buonanno:2022pgc}.

Adding classical spin to the Post-Minkowskian expansion leads to interesting challenges in the amplitude approach due to the traditional barrier at  spin-2 in quantum
field theory (although recent progress in describing massive higher spin states has been made in Refs. \cite{Bautista:2021wfy,
Chiodaroli:2021eug,
Cangemi:2022abk,Ochirov:2022nqz}). Results at the first Post-Minkowskian order and all orders in the spins were derived 
by solving Einstein's field equations directly \cite{Vines:2017hyw}. Amplitude-based and worldline approaches have since made substantial progress towards
obtaining Post-Minkowskian results with spin \cite{Guevara:2017csg,Vines:2018gqi,Guevara:2018wpp,Chung:2018kqs,Maybee:2019jus,Chung:2019duq,Damgaard:2019lfh,Aoude:2020onz,Bern:2020buy,Haddad:2020tvs,Kosmopoulos:2021zoq,Haddad:2021znf,Chen:2021qkk,Bern:2022kto,Alessio:2022kwv,delaCruz:2022nlj,FebresCordero:2022jts,Kim:2022iub,Guevara:2020xjx,
Arkani-Hamed:2019ymq,
Cristofoli:2021jas,Siemonsen:2019dsu,Bautista:2019tdr,Adamo:2021rfq,Menezes:2022tcs,Riva:2022fru}. In order to have known limits in which to compare amplitude-based results for scattering angles with those computed directly from general relativity, we here 
reconsider the classical problem of scattering in the equatorial plane of a Kerr black hole. We restrict the spin of the black hole to be parallel with the
orbital angular momentum and the motion is therefore restricted to the equatorial plane. A single scattering angle can then describe the asymptotic motion 
and the situation is rather similar to that of scattering around a Schwarzschild black hole except for the fact that the scattering angle
will depend on whether the black hole spin is pointing in the same direction as the orbital angular momentum, or opposite. We will be 
working with metrics of signature $(-,+,+,+)$ throughout.

One of the interesting observations of the first-order 
Post-Minkowskian result of ref.~\cite{Vines:2017hyw} was that the spins, to that order in the Post-Minkowskian expansion, could be provided in an exact
(resummed) form. The same resummed form naturally appears also from amplitude calculations to the same order in the Post-Minkowskian
expansion \cite{Guevara:2018wpp} and remnants of such a structure can be found also at second Post-Minkowskian order, at least up to fourth order
in one of the spins \cite{Guevara:2018wpp}. It turns out that this structure of resummed spin is a general feature of the probe limit: if the probe
is spinless, we can show this to any order in the Post-Minkowskian expansion. Taking the lightlike limit,
 and expanding in the black hole spin,
we recover the Kerr results for the bending of light \cite{Iyer:2009hq}. As we shall detail below, there are several other checks on our results as well.

Finally, an interesting and challenging problem is that of adding spin to the probe. 
We shall derive expressions for the Kerr scattering angle for a spinning probe, with the probe spin aligned with both the orbital angular momentum and the Kerr spin, valid up to (and including) second order in the black-hole
probe spin.
In principle, these calculations can be carried through to arbitary Post-Minkowskian order and we illustrate that below by providing analytical 
expressions up to and including ${\mathcal O}(G^5)$. Truncating to third Post-Minkowskian order our
results agree with those of refs. \cite{Vines:2018gqi,Jakobsen:2022fcj,FebresCordero:2022jts}.  The general expressions we present here for the probe
limit both without and with spin may be useful for checks on amplitude computations at higher orders in the Post-Minkowskian expansion.

\section{Warm-up: Scattering in Schwarzschild Metrics}

Before we turn to the main subject, it is instructive to describe our method in a far simpler setting that still retains the important features.
We therefore first consider the computationally easier problem of scattering around a Schwarzschild black hole. This will highlight the importance of
choosing suitable variables to simplify the calculation.

Consider first a scattering problem in a spherically symmetric effective potential $V_{\rm eff}(r)$ for which the radial momentum reads
\beq
p_r ~=~ \sqrt{p_{\infty}^2 - \frac{L^2}{r^2} - V_{\rm eff}(r)} ~ \label{radial_action_0},
\eeq
where $p_{\infty}$ is the three-momentum at radial infinity and $L$ is the conserved angular momentum. As is well known from analytical 
mechanics (say, fom Hamilton-Jacobi theory),
the scattering angle $\chi$ in such a situation follows from the relation 
\beq\label{echi}
\frac{\chi}{2} = -\frac{\partial}{\partial L}\int_{r_m}^{\infty}dr \sqrt{p_{\infty}^2 - \frac{L^2}{r^2} - V_{\rm eff}(r)} - \frac{\pi}{2},
\eeq
where $r_m$ is the turning point of the orbit. This is determined by the condition $p_r(r_m) = 0$, i.e. at the (real and positive) 
point where the integrand vanishes.
One may legitimately move the derivative with respect to $L$ inside the integral since the boundary term at $r_m$ vanishes by definition. The scattering angle
can thus be computed from
\beq
\frac{\chi}{2} = L \int_{r_m}^{\infty}\frac{dr}{r^2} \frac{1}{\sqrt{p_{\infty}^2 - \frac{L^2}{r^2} - V_{\rm eff}(r)}}  - \frac{\pi}{2} ~,
\eeq
which not only appears to depend on $r_m$ but even seems to be singular due to the integrand diverging at the endpoint. As is well known,
these subtleties are only apparent and the whole expression is completely well defined. In reality, though, except for a very small set of integrable potentials
$V_{\rm eff}(r)$, we wish to find the scattering angle as a perturbative series in the strength of the potential. A very compact solution to this
problem was recently provided in ref.~\cite{Bjerrum-Bohr:2019kec}, where the scattering angle is given in terms of a series of finite integrals,
with one new integral appearing for each order in perturbation theory. The final result reads
\beq
\chi = \sum_{k=1}^{\infty}
\frac{2b}{k!}
\int_0^{\infty}du \left(\frac{d}{du^2}\right)^k
\frac{[V_{\rm eff}(r)]^kr^{2(k-1)}}{p_{\infty}^{2k}} \label{chibasic} ~.
\eeq
Here $r^2 = u^2 + b^2$ and the impact parameter $b$ has been introduced in the usual way by $b = L/p_{\infty}$. Note that all integrals now run
along the full positive line, and they thus become elementary for power-law potentials. 
One important example which immediately fits into this framework is that of scattering in a Schwarzschild metric expressed in isotropic coordinates, and thus with line element
\beq
ds^2 = -\left(\frac{1 + \frac{GM}{2r}}{1 - \frac{GM}{2r}}\right)^2dt^2 + \left(1 + 
\frac{GM}{2r}\right)^4\left(dr^2 + r^2(d\theta^2 + \sin^2\theta d\phi^2)\right).
\eeq
This translates into the effective potential \cite{Damgaard:2021rnk}
\beq
V_{\rm eff}(r) = m^2(\gamma^2-1)
-m^2\left(1+{GM\over2r}\right)^4\left(\gamma^2\left(1+{GM\over2r}\over
  1-{GM\over2r}\right)^2-1\right) ~.
\eeq
Here $\gamma = 1/\sqrt{1-v^2}$ is the usual Lorentz contraction factor and we have
chosen the scattering to take place in the equatorial plane of $\theta = \pi/2$. 
Writing down the Schwarzschild scattering angle to any order in $G$ is thus straightforward upon expansion of the potential in a power series and subsequent use of eq.~(\ref{chibasic}).

We now wish to generalize the derivation of ref.~\cite{Bjerrum-Bohr:2019kec} so that it is amenable to more general metrics. We will
follow the standard approach based on solving for the radial momentum $p_r$. However, for general metrics, and in particular also for the 
Schwarzschild metric in coordinates different from isotropic, this will not lead to an expression of the simple form (\ref{radial_action_0}). In 
order to retain as many as possible of the simplifying features of the approach followed in ref.~\cite{Bjerrum-Bohr:2019kec} we will
make a suitable change of variables to a metric which in the limit of no interactions ($G \rightarrow 0$) reduces to the metric of flat Minkowski space in 
spherical coordinates. As a consequence, we recover the simple relation
\beq
p_r = \sqrt{p_{\infty}^2 - L^2/r^2}
\label{free-pr}
\eeq
in that limit. We will refer to metrics with
this property as
being in {\em normal form}. An example will best illustrate what we
mean. To this end, let us consider the Schwarzschild metric, but now written in standard Schwarzschild coordinates
\beq
ds^2 = -\left(1 + \frac{r_s}{r}\right)^2dt^2 + \left(1 - 
\frac{r_s}{r}\right)^{-1}dr^2 + r^2(d\theta^2 + \sin^2\theta d\phi^2),
\eeq
where we have defined $r_s \equiv 2GM$. For the obvious choice $\theta = \pi/2$ the metric leads to
\begin{eqnarray}
p_r^2 = \frac{(E^2-m^2)r^3+m^2r^2r_s-L^2(r-r_s)}{r(r-r_s)^2}.
\label{SWCpr}
\end{eqnarray}
In the free case, the radial momentum eq.~(\ref{SWCpr}) reduces to
$p_r^2 = p_{\infty}^2 - L^2/r^2$.
It is thus possible to separate out this part and write, equivalently,
\begin{eqnarray}
p_r^2=p_\infty^2-\frac{L^2}{r^2}-
r_s\left(\frac{m^2(r-r_s)-E^2(2r-r_s)+\frac{L^2}{r^2}(r-r_s)}{(r-r_s)^2}\right), \label{Schw_p_r}
\end{eqnarray}
where the last term involving the bracket clearly vanishes as $G \rightarrow 0$, and we have simply added and subtracted $L^2/r^2$. The advantage of this rewriting is that it makes it natural to interpret the remainder 
\beq
U(r,L) ~\equiv~
r_s\frac{m^2(r-r_s)-E^2(2r-r_s)+\frac{L^2}{r^2}(r-r_s)}{(r-r_s)^2} \label{SchwarzR}
\eeq
formally as an $L$-dependent potential. The Schwarzschild metric in these coordinates is therefore already of normal form, and the scattering angle can thus 
still be written as
\beq\label{eq:chi_deriv_rep}
\frac{\chi}{2} = -\frac{\partial}{\partial L}\int_{r_m}^{\infty}dr \sqrt{p_{\infty}^2 - \frac{L^2}{r^2} - U(r,L)} - \frac{\pi}{2},
\eeq
but the derivative will now also act on $U(r,L)$. We hence need to generalize the derivation of ref.~\cite{Bjerrum-Bohr:2019kec} to this new situation. 
Moreover, we discover that the $L$-derivative of eq.~(\ref{eq:chi_deriv_rep}), which is so natural from the canonical formalism, can be disposed of
so as to open up for more general situations including the spin of the probe.
Let us jump ahead to the final result which turns out to be surprisingly simple. In order to introduce it, we write the scattering angle in the form
\begin{equation}\label{eq:chiintdr}
  \frac{\chi}{2} =
  \int_{r_m}^\infty dr \frac{d\phi}{dr} - \frac{\pi}{2} = -\int_{r_m}^\infty \frac{h(r)}{p_\infty}\left(1-b^2/r^2-\frac{U(r,b)}{p_\infty^2}\right)^{-1/2}-
\frac{\pi}{2} ~,
  \end{equation}
where 
\beq
h(r) ~\equiv~ -\frac{d\phi}{dr}p_r.
\eeq
This rather trivial rewriting in fact anticipates, in simple cases, a first order derivative representation of $\chi$ in terms of
the radial action as in eq.~\eqref{eq:chi_deriv_rep}. Moreover, it allows for 
greater flexibility regarding the effective potentials we can handle. In terms of these quantities, the scattering angle is given by
\begin{equation}\label{eq_general}
\chi = -2\sum_{n=0}^\infty
\int_0^\infty du
\left(\frac{d}{du^2}\right)^{n}
h(r)
\frac{r^{2n}U(r,b)^n}{n!p_\infty^{2n+1}} - \pi, \quad \quad r^2=u^2+b^2 ~.
\end{equation}
The function $h(r)$ needs to be determined for each specific scattering situation but it often takes very simple forms. As an example, for the Schwarzschild metric in isotropic coordinates it is simply $h(r)=-bp_\infty/r^2$. The identification of $h(r)$ is useful for both non-spinning and spinning probes but it is particularly suited for the latter, where there may be no obvious way in which to relate the integrand of the scattering angle to a first-order derivative. As detailed in our derivation below, the formula (\ref{eq_general}) is valid for any $h(r)$ which is real analytic on the interval $r\in [r_m,\infty[$, and 
falls off as $\lim_{r\rightarrow\infty}h(r)\sim 1/r^{n}$, with $n\geq 2$. These conditions are always met for the cases considered in this paper.

\subsection{A compact formula for the scattering angle in metrics of normal form}

Although the final result eq.~(\ref{eq_general}) is surprisingly simple, the steps leading to it are involved and
we display them now with a fair amount of detail. 
Let us first introduce some general notation.
For any non-trivial metric, $p_r$ will depend on $G$ and this dependence carries all of the information about the scattering angle. We define
$T \equiv \left.p_r^2\right|_{G\rightarrow0}$ so that we can write
\begin{equation}\label{eq:prTR}
p_r^2=T(r)-U(r)
\end{equation}
where, by construction, $U(r)$ carries all the $G$-dependence. Both $T$ and $U$ depend on the radial coordinate $r$ as indicated but may also in general  
depend on orbital angular momentum $L$ and any other parameters of the metric. The function $U$ is a close analogue of a classical effective potential associated with the given metric (for some choice of coordinates). If $U$ carries no $L$-dependence the method used in ref.~\cite{Bjerrum-Bohr:2019kec} can straightforwardly be used to derive the scattering angle in perturbation theory.  Here we consider its generalization to the $L$-dependent setting, focusing on
a formulation that will encompass the case of spinning probes.

After having introduced this notation, we now return to the case of a metric which we assume is already in normal form. As explained
above, this means that $T(r)$ takes the simple form
\beq
T(r) ~=~ p_{\infty}^2 - \frac{L^2}{r^2}
\label{T-normal}
\eeq
in those coordinates.
We recall that we can then write the scattering angle as
\begin{equation}\label{eq:chiintdr}
  \frac{\chi}{2}=
  \int_{r_m}^\infty dr \frac{d\phi}{dr} - \frac{\pi}{2} = 
-\int_{r_m}^\infty \frac{h(r)}{p_\infty}\left(1-b^2/r^2-\frac{U(r,b)}{p_\infty^2}\right)^{-1/2}- \frac{\pi}{2}, \quad\quad \frac{d\phi}{dr}=-\frac{h(r)}{p_r},
  \end{equation}
with $h(r)$ assumed to obey the analyticity and fall-off requirements listed above. This will be the starting point for our derivation. 

Now, using the condition $p_r(r_m)=0$, and following the derivation of ref.~\cite{Bjerrum-Bohr:2019kec}, we find it convenient to isolate
  \begin{equation}\label{eq:rmcondition}
 \frac{ b^2}{r^2}
  =
  \frac{r_m^2}{r^2}-\frac{r_m^2}{r^2}\frac{U(r_m,b)}{p_\infty^2},
  \end{equation}
 and insert this into eq.~(\ref{eq:chiintdr}). This gives
  \begin{equation}
  \chi/2
  =
  -\int_{r_m}^\infty\!dr\, \frac{h(r)}{p_\infty}\left(1-\frac{r_m^2}{r^2}-W(r,b)\right)^{-1/2}-\pi/2,
  \end{equation}
where
\begin{equation}
W(r,b) \equiv \frac{1}{p_\infty^2}
\left(
U(r,b)-\frac{r_m^2}{r^2}U(r_m,b)
\right) .
\end{equation}
Changing integration variable to $u$ through $r^2=u^2+r_m^2$ (where $u \geq 0$), we get
\begin{equation}\label{eq:chiWexpready}
\frac{\chi}{2}
=
-\int_0^\infty\!du\, \frac{h(r)}{p_\infty}\left(1-\frac{r^2}{u^2}W(r,b)\right)^{-1/2}- \frac{\pi}{2},
\end{equation}
where $r$ just stands for $r=\sqrt{u^2+r_m^2}$. Use of the binomial expansion
\begin{equation}
(1+x)^{-1/2}=1+\sum_{n=0}^\infty \begin{pmatrix} -1/2\\n+1 \end{pmatrix}x^{n+1} = 1 + \sum_{n=0}^\infty  \frac{(-1)^{n+1}(2n+1)!!}{2^{n+1}(n+1)!} x^{n+1},
\end{equation}
yields the following expression for the angle
\begin{equation}\label{eq:chiF0sumform}
\frac{\chi}{2}
=
F_0(r_m)
-
\sum_{n=0}^\infty \frac{(2n+1)!!}{2^{n+1}(n+1)!}
\int_0^\infty\!du\,
\frac{1}{u^{2(n+1)}}\left(\frac{h(r)}{p_\infty}r^{2(n+1)}W(r,b)^{n+1}\right)
- \frac{\pi}{2},
\end{equation}
where we have defined the function 
\beq
F_0(r_m)\equiv -\frac{1}{p_{\infty}}\int_0^\infty du~ h(r) \quad\quad r^2 = u^2 + r_m^2 ~. 
\eeq
Although this integral is often elementary (such as for the Schwarzschild metric in isotropic coordinates), 
we do not need to evaluate it explicitly. This will
become clear below. In fact, this function, being dependent on $r_m$ must disappear in the end since the scattering angle should not depend on $r_m$.
The remaining terms above can be rewritten by means of the integration-by-parts identity \cite{Wallace:1973iu},
\begin{equation}\label{eq:wallaceidentity}
\int_0^\infty \frac{du}{u^{2(n+1)}}f(u)
=
\frac{1}{(2n+1)!!}\int_0^\infty du\left(\frac{1}{u}\frac{d}{du}\right)^{n+1} f(u)
=
\frac{2^{n+1}}{(2n+1)!!}\int_0^\infty du\left(\frac{d}{du^2}\right)^{n+1} f(u),
\end{equation}
valid for any $\mathcal C^\infty$ function $f$ for which $f(u)/u^{2n+1}$ vanishes at zero and infinity. 
On account of our assumptions about $h(r)$, eq.~\eqref{eq:wallaceidentity}  may be applied to eq.~\eqref{eq:chiF0sumform} to obtain 
\begin{eqnarray}\label{eq:chiDelta}
\frac{\chi}{2} &=& 
F_0(r_m)
-\sum_{n=0}^\infty \Delta_n  - \frac{\pi}{2},
\end{eqnarray}
where we have defined
\beq
\Delta_n ~\equiv~  \frac{1}{(n+1)!}
\int_0^\infty\!du
\left(\frac{d}{du^2}\right)^{n+1}
\left[\frac{h(r)}{p_\infty}r^{2(n+1)}W(r,b)^{n+1}\right].
\eeq
Furthermore, writing
\begin{equation}\label{eq:rewriteDbegin}
W(r,b)^{n+1}=\frac{U(r,b)^{n+1}}{p_\infty^{2(n+1)}}(1-x)^{n+1}\quad\text{ with }\quad x\equiv -\frac{r_m^2}{r^2}\frac{U(r_m,b)}{U(r,b)},
\end{equation}
we can again Taylor expand, this time in powers of $x$, to get
\begin{equation}
\Delta_n =
\int_0^\infty du \left(\frac{d}{du^2}\right)^{n+1}\,\,
\sum_{k=0}^{n+1}
\frac{1}{(n-k+1)!k!}
\frac{h(r)}{p_\infty}
\frac{r^{2(n+1)}U(r,b)^{n-k+1}}{p_\infty^{2(n+1)}}
\left[-\frac{r_m^{2}}{r^{2}}U(r_m,b)\right]^k,
\end{equation}
and once again we can use eq.~(\ref{eq:rmcondition}) to substitute the $U(r_m,b)$ in the square brackets. This results in
\begin{equation}
\label{Deltasum}
\Delta_n =
\sum_{k=0}^{n+1}
\frac{(b^2-r_m^2)^k}{k!}
\int_0^\infty du
\left(\frac{d}{du^2}\right)^{n+1}\,\,
\frac{h(r)}{p_\infty}
\frac{r^{2(n-k+1)}U(r,b)^{n-k+1}}{(n-k+1)!\,p_\infty^{2(n-k+1)}}.
\end{equation}
Note that the only explicit $r_m$-dependence in the integrand occurs through $r=\sqrt{u^2+r_m^2}$. Since $r$ is symmetric in $r_m^2$ and $u^2$, we can exchange derivatives in $u^2$ for derivatives in $r_m^2$, and consider the identity 
\beq
\left(\frac{d}{du^2}\right)^{n+1}
=
\left(\frac{d}{dr_m^2}\right)^{k}\left(\frac{d}{du^2}\right)^{n-k+1} ~.
\eeq
Applying this to the sum in eq.~(\ref{Deltasum}), we find 
\beq
\Delta_n=\sum_{k=0}^{n+1}\Delta_{n,k}(r_m) ~,
\eeq 
where we have defined
\begin{equation}\label{eq:deltank}
    \Delta_{n,k}\equiv\frac{(b^2-r_m^2)^k}{k!}
    \left(\frac{d}{dr_m^2}\right)^k
    \int_0^\infty \!du\,
    \left(\frac{d}{du^2}\right)^{n-k+1}\,\frac{h(r)}{p_\infty}
    \frac{r^{2(n-k+1)}U(r,b)^{n-k+1}}{(n-k+1)!\,p_\infty^{2(n-k+1)}} ~.
\end{equation}
We observe that the term with $k=n+1$ is $U$-independent. Crucially, as we shall demonstrate next, this fact 
will make the apparent $r_m$-dependence disappear, cancelling the $r_m$-dependent piece $F_0(r_m)$.
We start by evaluating the $k=n+1$ and $F_0(r_m)$ terms together, and introduce (the reason for the factor 1/2 on 
the left hand side will become 
clear shortly),
\begin{equation}
\frac{1}{2}\zeta_{-1}\equiv F_0(r_m)-\sum_{n=0}^\infty \Delta_{n,n+1}(r_m).
\label{zetam}
\end{equation}
Consider now the Taylor expansion of $F_0(r_m)$ around $r_m=b$. This reads
\begin{equation}
F_0(b)=-\sum_{n=0}^\infty
    \frac{(b^2-r_m^2)^{n}}{n!}
    \left(\frac{d}{dr_m^2}\right)^{n}
    \int_0^\infty \!du\,
    \,\frac{h(r)}{p_\infty}.
    \label{TaylorF0b}
\end{equation}
Furthermore, we note that the sum $\sum_{n=0}^\infty \Delta_{n,n+1}$ can be rewritten as
\begin{equation}
\begin{split}
    \sum_{n=0}^\infty\Delta_{n,n+1}
    &=
    \sum_{n=0}^\infty
    \frac{(b^2-r_m^2)^{n+1}}{(n+1)!}
    \left(\frac{d}{dr_m^2}\right)^{n+1}
    \int_0^\infty \!du\,
    \,\frac{h(r)}{p_\infty},\\
    &=
    -
    \int_0^\infty du \frac{h(r)}{p_\infty}
    +
    \sum_{n=0}^\infty
    \frac{(b^2-r_m^2)^{n}}{n!}
    \left(\frac{d}{dr_m^2}\right)^{n}
    \int_0^\infty \!du\,
    \,\frac{h(r)}{p_\infty},
\end{split}
\end{equation}
where in the second line we have added and subtracted $F_0(r_m)$. Making use of eq.~(\ref{TaylorF0b}) and the definition of $F_0(r_m)$ we find 
\beq
\sum_{n=0}^\infty\Delta_{n,n+1}=F_0(r_m)-F_0(b) ~.
\eeq
 Inserting this into eq.~(\ref{zetam}) results in $\frac{1}{2}\zeta_{-1}=F_0(b)$. The $r_m$-dependence has explicitly disappeared from this term. It follows from the above that the scattering angle can be written in the form
\beq
\chi-\zeta_{-1}+\pi = 
-2\sum_{n=0}^\infty\sum_{k=0}^n
\frac{(b^2-r_m^2)^k}{k!}
\left(\frac{d}{dr_m^2}\right)^k
\int_0^\infty du
\left(\frac{d}{du^2}\right)^{n-k+1}\,\,
\frac{h(r)}{p_\infty}
\frac{r^{2(n-k+1)}U(r,b)^{n-k+1}}{(n-k+1)!\,p_\infty^{2(n-k+1)}}
\eeq
To simplify our notation, we now define
\begin{equation}\label{eq:zetanx}
\zeta_n(x) \equiv -2\int_0^\infty du \left(\frac{d}{du^2}\right)^{n+1}
\frac{h(r)}{p_\infty}
\frac{r^{2(n+1)}U(r,b)^{n+1}}{(n+1)!p_\infty^{2(n+1)}}, \quad\quad r^2=u^2+x^2
\end{equation}
so that
\begin{eqnarray}
  \chi-\zeta_{-1}+\pi &=&
\sum_{n=0}^\infty\sum_{k=0}^n
\frac{(b^2-r_m^2)^k}{k!}
\left(\frac{d}{dr_m^2}\right)^k
\zeta_{n-k}(r_m) \cr
&=& \sum_{n=0}^\infty\sum_{\ell=0}^\infty
\frac{(b^2-r_m^2)^{\ell}}{\ell!}
\left(\frac{d}{dr_m^2}\right)^{\ell}
\zeta_{n}(r_m)
\end{eqnarray}
which we recognize as the Taylor expansion of $\zeta_{n}(b)$ around the point $r_m$. 
In this way, the turning point $r_m$ has explicitly disappeared from all terms of
the scattering angle, as it should. No regularization of the involved integrals and no use of {\em ad hoc} rules has been needed. The final formula for the scattering angle thus becomes
\begin{equation}
\chi +\pi=
\sum_{n=0}^\infty \zeta_n(b) +\zeta_{-1},
\label{Jitzeeq_almost}
\end{equation}
where $\zeta_n(b)$ is given in eq.~(\ref{eq:zetanx}) evaluated, as we see, at $r^2=u^2+b^2$.
The choice of notation for $\zeta_{-1}$ is now clear, as this is precisely $\zeta_n$ from eq.~\eqref{eq:zetanx} evaluated at $x=b$ and $n=-1$. 
Thus we can write eq.~\eqref{Jitzeeq_almost} as 
\begin{equation}\label{eq_general}
\chi =-2\sum_{n=0}^\infty
\int_0^\infty
du
\left(\frac{d}{du^2}\right)^{n}
h(r)
\frac{r^{2n}U(r,b)^n}{n!p_\infty^{2n+1}} - \pi, \quad\quad r^2=u^2+b^2
\end{equation}
This is a very general result, valid for any well-behaved $h(r)$ as stipulated in precise terms above. As we shall see next, it will apply to
both scalar and spinning test bodies 
up to cubic order in the spin of the test particle $S$ and we see no obstacle towards it being applicable to any order in the spin of the probes.
The essential ingredient is that the spin of the test body is considered in perturbation theory. Note that when interactions are turned off,
the function $h(r)$ reduces to $h(r) = -bp_{\infty}/r^2$ and the $n=0$ term in the sum above thus produces zero scattering angle up to terms
that vanish when interactions are set to zero. 

Having derived this general expression for the scattering angle, it is of interest to see the form it takes when the scattering angle
can be expressed in terms of an $L$-derivative of the radial action as in eq.~(\ref{eq:chi_deriv_rep}). We will then also see how 
it relates to the formula derived in ref.~\cite{Bjerrum-Bohr:2019kec} for the special case where $U(r,b)$ does not depend on $b$.
Carrying out the derivative in eq.~(\ref{eq:chi_deriv_rep}), we see that
\beq
  \frac{d\phi}{dr}=-\frac{\partial}{\partial L}p_r=\frac{1}{p_r}\left(\frac{bp_\infty}{r^2}+\frac{1}{2p_\infty}
\frac{\partial}{\partial b} U(r,b)\right),
\eeq
and thus
\beq
h(r)=-\left(\frac{bp_\infty}{r^2}+\frac{1}{2p_\infty}\frac{\partial}{\partial b} U(r,b)\right) ~.
\label{special-h}
\eeq
Let us define 
\beq
\Upsilon(r,b) \equiv \frac{\partial}{\partial b}U(r,b) ~.
\eeq
Inserting $h(r)$ from eq.~(\ref{special-h}) into \eqref{eq_general}, we get
\begin{equation}\label{eq:chirewrittenalt}
    \begin{split}
    \chi &= 2\sum_{n=0}^\infty\int_0^\infty \!du\, \left(\frac{d}{du^2}\right)^{n}\left[\frac{bp_\infty}{r^2}+\frac{1}{2p_\infty}\Upsilon(r,b)\right]
\frac{r^{2n}U(r,b)^n}{n!p_\infty^{2(n+1)-1}} - \pi\\
    &=2b\sum_{n=0}^\infty\int_0^\infty \!du\, \left(\frac{d}{du^2}\right)^{n+1}\frac{r^{2n}U(r,b)^{n+1}}{(n+1)!p_\infty^{2(n+1)}}+\sum_{n=0}^\infty\int_0^\infty \!du\, \left(\frac{d}{du^2}\right)^{n}\Upsilon(r,b)\frac{r^{2n}U(r,b)^n}{n!p_\infty^{2(n+1)}}
\end{split}
\end{equation}
where we recall the notation $r^2 = u^2 + b^2$. Note that the $n=0$ term from the free part has cancelled the explicit $\pi$, and we 
have relabelled the remaining terms accordingly. This is a valid and compact form for the scattering angle but we can simplify it
further by use of the identity (which is valid for $r^2 = u^2 + b^2$),
\beq
\frac{d}{db}\left[r^{2n}U(r,b)^{n+1}\right] = (n+1)r^{2n}\Upsilon(r,b)U(r,b)^n + 2b\frac{d}{d u^2}\left[r^{2n}U(r,b)^{n+1}\right] ~.
\eeq
Substituting this into eq.~(\ref{eq:chirewrittenalt}) we obtain the compact expression
\begin{equation}
    \chi=
    \sum_{n=0}^\infty\int_0^\infty \!du\, \left(\frac{d}{du^2}\right)^{n}\frac{d}{db}
\left[\frac{r^{2n}U(r,b)^{n+1}}{(n+1)!p_\infty^{2(n+1)}}\right], \quad\quad r^2=u^2+b^2
\label{L-derivativeangle}
\end{equation}
In the special case of a $b$-independent $U(r,b)$ this is seen to reduce to the formula (\ref{chibasic}).
Remarkably, whether based on the general formula (\ref{eq_general}) or on the special case (\ref{L-derivativeangle}),
we have found that these final results for the scattering angle are almost as simple as those given in ref.~\cite{Bjerrum-Bohr:2019kec}
even though $U(r,b)$, which effectively acts as a potential, here can depend on angular momentum $L$ (or $b=L/p_{\infty}$). The
more general representation (\ref{eq_general}) is valid
even when we cannot obviously write the scattering angle as an $L$-derivative of the radial action.

If we substitute the specific form of
$U(r,L)$ for Schwarzschild coordinates as in eq.~(\ref{SchwarzR}) we do indeed recover the correct Schwarzschild scattering angles from
eq.~(\ref{eq_general}). To illustrate this point, we perform the Taylor expansion of the $U(r,b)$ in eq.~(\ref{SchwarzR}) to second order in $G$. This results in
\begin{equation}
U(r,b) = U_1(r,b) + U_2(r,b) + \mathcal O(G^3) = -\frac{(2E^2-m^2)r^2-L^2}{r^3}r_s - \frac{(3E^2-m^2)r^2-L^2}{r^4}r_s^2 + \mathcal O(G^3) ~.
\end{equation}
To leading order in $G$ only the single term $U_1(r,b)$ contributes to the scattering angle and we thus get
\begin{equation}
\chi_1
=
\int_0^\infty du \frac{d}{d b}\frac{1}{p_\infty^2}U_1(r,b)
=
\frac{2GM(2E^2-m^2)}{bp_\infty^2}
=
\frac{2(2\gamma^2-1)GMm^2}{bp_\infty^2},
\end{equation}
where $E^2=\gamma^2m^2$ and $\gamma = 1/\sqrt{1-v^2}$ is the Lorentz factor of the test-particle with velocity $v$ at infinity. This is the well-known leading order result.
At second order in $G$ there are two contributions: the $U_1$ term from $n=1$ in the sum (\ref{L-derivativeangle}) 
and the $U_2$ term from $n=0$ part.
The second-order contribution $\chi_2$ to the scattering angle is thus 
\begin{align}
\nonumber
\chi_2 =
\int_0^\infty du \frac{\partial}{\partial b} \left(\frac{d}{du^2}\right)\frac{r^2 U^2_1(r,b)}{2p_\infty^{4}}  +
\int_0^\infty du \frac{\partial}{\partial b}
\frac{U_2(r,b)}{p_\infty^{2}} &=
r_s^2\left(\frac{\pi(6E^2-2m^2-p_\infty^2)}{4p_\infty^2b^2}
-
\frac{\pi(8E^2-4m^2-3p_\infty^2)}{16b^2p_\infty^2}
\right) \\
=
4G^2M^2m^2\pi\left(\frac{5\gamma^2-1}{4p_\infty^2b^2}-\frac{5\gamma^2-1}{16p_\infty^2b^2}\right)
&=
\frac{3G^2M^2m^2\pi(5\gamma^2-1)}{4p_\infty^2b^2}
\end{align}
which is the known answer. In Table 1 below we list the contributions up to and including tenth order in $G$ computed straightforwardly in this manner, but expressed in terms of velocity $v$ rather than $\gamma$ for the sake of compactness.
  \def\arraystretch{1.5}
\begin{table}[h!]
  \begin{tabular}{| c | p{0.60\paperwidth} |}
  \hline
  $n$ & $\chi_n/\frac{G^nM^n}{b^nv^{2n}}$\\\hline\hline
  1 &  $2\left(v^2+1\right)$\\\hline
  2 &  $(3 \pi/4)v^2\left(v^2+4\right)$\\\hline
  3 & $(2/3)  \left(5 v^6+45 v^4+15 v^2-1\right)$ \\\hline
  4 & $(105\pi/64) v^4\left(v^4+16 v^2+16\right)$ \\\hline
  5 & $(2/5) \left(21 v^{10}+525 v^8+1050 v^6+210 v^4-15 v^2+1\right)$\\\hline
  6 & $(1155 \pi/256) v^6\left(v^6+36 v^4+120 v^2+64\right)$\\\hline
  7 & $(2/35) \left(429 v^{14}+21021 v^{12}+105105 v^{10}+105105 v^8+15015 v^6-1001 v^4+91 v^2-5\right)$\\\hline
  8 & $(45045 \pi/16384)   v^8\left(5 v^8+320 v^6+2240 v^4+3584 v^2+1280\right)$\\\hline
  9 & $(2/63)  (2431 v^{18}+196911 v^{16}+1837836 v^{14}+4288284 v^{12}
$\newline
$  
+2756754 v^{10}+306306 v^8-18564 v^6+1836 v^4-153 v^2+7)$\\\hline
  10 & $(2909907\pi/65536)  v^{10}\left(v^{10}+100 v^8+1200 v^6+3840 v^4+3840 v^2+1024\right)$\\\hline
  \end{tabular}
\caption{Scattering angle of a non-spinning test particle in a Schwarzschild background up to 10th order in $G$. \\$\chi_n$ is the $G^n$ contribution to the full scattering angle $\chi=\sum_n\chi_n$.}
\end{table}


\def\arraystretch{1.5}
\begin{table}[h!]
\begin{tabular}{|c|p{0.93\linewidth}|}
\hline
$n$ & $\chi_n/\frac{G^nM^n}{v^{2n}(b^2-a^2)^{(3n-1)/2}}$\\ 
\hline\hline
1 & $2 \big[-2 a v+b(1+v^2)\big]$\\
\hline\hline
2 & $(\pi/2a^2) 
\Big[
(b^2-a^2)^{5/2}v^4+(a-bv)^3
\big[-4a^2v+3ab+b^2 v\big]
\Big]$
\\\hline\hline
3 & $(2/3) \big[
2 a^5 v \left(3 v^4-10
   v^2-9\right)-3 a^4 b
   \left(v^6-15 v^4-45
   v^2-5\right)-4 a^3 b^2 v
   \left(15 v^4+70
   v^2+27\right)
$\newline$
+2 a^2 b^3
   \left(11 v^6+135 v^4+105
   v^2+5\right)-18 a b^4 v
   \left(5 v^4+10
   v^2+1\right)+b^5 \left(5
   v^6+45 v^4+15 v^2-1\right)
\big]$
\\\hline\hline
4 & $(3 \pi/16a^4) 
\Big[
2(b^2-a^2)^{11/2}v^8+(a-bv)^5\big[-8 a^6 v \left(14 v^2+5\right)+5
   a^5 b \left(72
   v^2+7\right)
$\newline$
+a^4 b^2 v
   \left(16 v^2-305\right)
-5 a^3
   b^3 \left(11
   v^2-14\right)-a^2 b^4 v
   \left(11 v^2-30\right)+10 a
   b^5 v^2+2 b^6 v^3\big]\Big]$
 \\\hline\hline
 5 & $(2/15) \big[
-2 a^9 v \left(15 v^8-60 v^6+378
   v^4+900 v^2+175\right)
$\newline$   
   +15 a^8
   b \left(v^{10}-15 v^8+210
   v^6+1050 v^4+525
   v^2+21\right)
+8 a^7 b^2 v
   \left(45 v^8-780 v^6-6426
   v^4-6300 v^2-875\right)
$\newline$   
   -140
   a^6 b^3 \left(v^{10}-45
   v^8-630 v^6-1050 v^4-315
   v^2-9\right)
-84 a^5 b^4 v
   \left(45 v^8+1020 v^6+2814
   v^4+1500 v^2+125\right)
$\newline$   
   +14
   a^4 b^5 \left(67 v^{10}+3375
   v^8+15750 v^6+14070 v^4+2295
   v^2+27\right)
-1400 a^3 b^6 v
   \left(9 v^8+84 v^6+126 v^4+36
   v^2+1\right)
$\newline$   
   +36 a^2 b^7
   \left(29 v^{10}+875 v^8+2450
   v^6+1190 v^4+65
   v^2-1\right)
-50 a b^8 v
   \left(63 v^8+420 v^6+378
   v^4+36 v^2-1\right)
$\newline$   
   +3 b^9
   \left(21 v^{10}+525 v^8+1050
   v^6+210 v^4-15 v^2+1\right)
\big]$
\\\hline\hline
6 & $(5 \pi/128a^6)
\Big[
8(b^2-a^2)^{17/2}v^{12}+(a-bv)^7\big[-4 a^{10} v \left(1584 v^4+1540
   v^2+189\right)
$\newline$
+7 a^9 b
   \left(5720 v^4+2808
   v^2+99\right)
-a^8 b^2 v
   \left(2200 v^4+85232
   v^2+17829\right)+14 a^7 b^3
   \left(260 v^4+5391
   v^2+330\right)
$\newline$
-2 a^6 b^4 v
   \left(334 v^4-1491
   v^2+14070\right)
+21 a^5 b^5
   \left(85 v^4-272
   v^2+176\right)+a^4 b^6 v
   \left(255 v^4-1904
   v^2+1680\right)
$\newline$
-28 a^3 b^7
   v^2 \left(17 v^2-24\right)
-4
   a^2 b^8 v^3 \left(17
   v^2-56\right)+56 a b^9 v^4+8
   b^{10} v^5
   \big]
\Big]$
\\ \hline
\end{tabular}

\caption{Scattering angle of a non-spinning test particle in a Kerr background up to 6th order in $G$. \\$\chi_n$ is the $G^n$ contribution to the full scattering angle $\chi=\sum_n\chi_n$.}
\label{table:kerrscattering}
\end{table}


\section{Scattering in Kerr Metrics}

We next consider applying the formula we found in the previous section to the scattering of a small non-spinning probe around a Kerr black hole.
A standard choice for the metric is Boyer-Lindquist coordinates $(t,r,\phi)$, for which, when restricted to the equatorial $\theta=\pi/2$ plane, the metric reads
\begin{equation}\label{eq:BLthetapi2}
g_{\mu\nu}
=
\begin{pmatrix}
-\left(1-\frac{r_s}{r}\right) & 0 & -\frac{r_sa}{r}\\
0 & \frac{r^2}{r^2-r_sr+a^2} & 0\\
-\frac{r_sa}{r} & 0 & \frac{(r+r_s)a^2+r^3}{r}
\end{pmatrix}.
\end{equation}
Letting a test body orbit in this $\theta=\pi/2$ plane, it will have its orbital angular momentum $L$ conserved, and it will  therefore remain 
in that plane. This allows for a well-defined scattering angle and it will also allow us to rewrite the Kerr metric in normal form.
We find the radial momentum $p_r$ from the Hamilton-Jacobi equation
\begin{equation}
p_r^2 =
\frac{r\left(p_\infty^2r^3+m^3r^2r_s+(a^2p_\infty^2-L^2)r+r_s(Ea-L)^2\right)}{(a^2+r^2-r_sr)^2},
\end{equation}
and we can write it in the form $p_r^2 = T(r,L,a) - U(r,L,a)$, with
\begin{equation}
T(r,L,a)  \equiv \frac{r^2}{r^2+a^2}\left(p_\infty^2-\frac{L^2}{r^2+a^2}\right),
\end{equation}
which indeed is independent of $G$, and 
\begin{eqnarray}\label{eq:RBL}
U(r,L,a)  &\equiv&
-\frac{
\left[(2E^2-m^2)r^6+(m^2-E^2)r^5r_s+((4E^2-m^2)a^4-4ELa^3)r^2\right]rr_s}{(a^2+r^2-r_sr)^2(a^2+r^2)^2} \cr
&-& \frac{\left[((5E^2-2m^2)a^2-2ELa-L^2)r^4-r_s((E^2-m^2)a^2-L^2)r^3+a^4(Ea-L)^2
\right]rr_s}{(a^2+r^2-r_sr)^2(a^2+r^2)^2},
\end{eqnarray}
which carries all $G$-dependence.
Although well separated into $T$ and $U$ pieces, we notice that $T$  is not of the free kind shown in eq.~(\ref{T-normal}). Thus, Boyer-Lindquist coordinates are not of
normal form and we need to choose different coordinates in order for our formalism to be applicable. As noted in the previous section, the needed change of 
integration variables in the radial action can equivalently be viewed as a coordinate transformation away from Boyer-Lindquist coordinates, thus leading to a
different metric.

Indeed, in the $G \to 0$ limit the Boyer-Lindquist metric eq.~(\ref{eq:BLthetapi2}) takes the form
\begin{equation}
g_{\mu\nu}
=
\begin{pmatrix}
-1 & 0 & 0\\
0 & \frac{r^2}{r^2+a^2} & 0\\
0 & 0 & a^2+r^2
\end{pmatrix}
\end{equation}
which does not correspond to flat Minkowski space in ordinary polar coordinates.
Since the Kerr metric is diagonal and only depends on the radial coordinate $r$ in this limit, we can find a coordinate change  $r \rightarrow \rho(r)$ which 
allows us to recover the free structure of $T$. This change is given simply by
\begin{equation}
\rho^2=r^2+a^2 ~. 
\label{Kerr-normal}
\end{equation}
For this new radial coordinate $\rho$ the Kerr metric takes the form
\begin{equation}
\tilde g_{\mu\nu} = \begin{pmatrix}
-1 & 0 & 0\\
0 & 1 & 0\\
0 & 0 & \rho^2
\end{pmatrix}
\end{equation}
in the $G \to 0$ limit, corresponding to a metric in normal form.
The transformation eq.~(\ref{Kerr-normal}),   also automatically produces the needed 
\beq
T(\rho,L,a) = p_{\infty}^2 - \frac{L^2}{r^2} ~.
\eeq
Note that the free part of the Kerr metric becomes
independent of the black-hole spin $a$ in these coordinates. The radial momentum $p_r$ transforms like
\begin{equation}\label{eq:prtrans}
p_\rho=\frac{dr}{d\rho} p_r
\end{equation}
under this coordinate change, and so we obtain the new effective potential
\beq
\tilde U(\rho,b,a)=\left(\frac{dr}{d\rho}\right)^2U(r,b,a).
\eeq
We may thus write the formula for the scattering angle (\ref{L-derivativeangle}) in terms of $\rho$ as
\begin{equation}
\chi
=
\sum_{n=0}^\infty
\int_0^\infty
du
\frac{\partial}{\partial b}
\left(\frac{d}{du^2}\right)^n
\frac{\rho^{2n}\tilde U(\rho,b)^{n+1}}{(n+1)!p_\infty^{2(n+1)}}, \quad \quad \rho^2=u^2+b^2. \label{chikerr}
\end{equation}
As a first quick check, we compute the scattering angle $\chi_1$ to leading order in $G$ using eq.~(\ref{chikerr}). We find
\begin{equation}\label{echikerr1PM}
\chi_1
=
\frac{2 G M \left(\gamma ^2 (2 b-2 a v)-b\right)}{\gamma ^2 v^2 \left(b^2-a^2\right)}
\end{equation}
where $v$ is the asymptotic velocity of the test-particle. This agrees with the scattering angle computed in ref.~\cite{Vines:2017hyw} when restricted to the test-body limit. We emphasize that as
in ref.~\cite{Vines:2017hyw} our result gives
the scattering angle to all orders in the spin of the black hole $a$. 
It is easy to verify that in the light-like limit $v \to 1, \gamma \to \infty$,  
the result above reproduces the terms of the expansion provided in ref.~\cite{Iyer:2009hq}.

Although the integrals are slightly more involved than those of Schwarzschild scattering, the final results for massive-probe scattering are relatively simple. 
In table \ref{table:kerrscattering} we list results up to and including sixth order in $G$ (this table can readily be extended based on our general formula). Again, expanding
in powers of $a$ and taking the massless limit this reproduces the well-known light-bending formulas for Kerr metrics.

We note that the resulting scattering angle contribution $\chi_n$ to any order in $G$ displays some simple patterns. First, the scattering angle naturally has emerged in
a form that re-sums all orders in $a$. Second, to order $n$ one may identify the prefactor
\begin{equation}
c_n\equiv \frac{G^nM^n}{(b^2 - a^2)^{(3n-1)/2}v^{2n}}.
\label{spinlessPF}
\end{equation}
which accounts for the all-order-in-spin behaviour. 
We also note that even orders in powers of $G$ are relatively simpler, and one can easily identify more structural patterns in them. To be concrete, we observe that after factorising the term in eq.~(\ref{spinlessPF}), the angle takes the form
\begin{equation}
\chi_n/c_n=(d_{1,n}\pi/a^{2n})\Big[
d_{2,n}v^{2n}(b^2 - a^2)^{(3n-1)/2}
+(a-bv)^{n+1}\sum_{\ell=0}^{2n-2}a^\ell b^{2n-2-\ell}f_{n,\ell}(v)
\Big]
\end{equation} 
where $d_{i,n}$ are numerical constants, and $f_{n,k}(v)$ is a polynomial in $v$ of degree $n-1$ for even $\ell$ or $n-2$ for odd $\ell$.

\def\arraystretch{1.5}
\begin{table}[h!]
\begin{tabular}{|c|p{0.80\linewidth}|}
\hline
$(n,k)$ & $\chi_{n,k}/\frac{G^nM^n(S/m)^k}{v^{2n}(b^2-a^2)^{(3n+2k-1)/2}}$\\
\hline\hline
$(1,1)$
 & 
$
-4 (a v-b) (a-b v)
$
\\ \hline
$(1,2)$
&
$
-4 a^3 v+6 a^2 b
   \left(v^2+1\right)-12 a b^2
   v+2 b^3 \left(v^2+1\right)
$
\\\hline\hline
$(2,1)$ &
$
(3 \pi/2)(a v-b) (a-b v)\big[
a^2 \left(-2 v^2-3\right)+10 a b
   v-b^2 \left(3 v^2+2\right)
\big]
$
\\ \hline
$(2,2)$
&
$
(3 \pi/4) \big[
-10 a^5 v \left(v^2+1\right)
+a^4
   b \left(12 v^4+71
   v^2+12\right)
   -90 a^3 b^2 v
   \left(v^2+1\right)
   $\newline$
   +a^2 b^3
   \left(21 v^4+128
   v^2+21\right)-40 a b^4 v
   \left(v^2+1\right)+b^5
   \left(2 v^4+11 v^2+2\right)
\big]
$
\\\hline\hline
$(3,1)$ &
$8(a v-b) (a-b v)\big[
a^4 \left(-v^4-10 v^2-5\right)+8
   a^3 b v \left(3
   v^2+5\right)
$\newline$   
   -2 a^2 b^2
   \left(5 v^4+38 v^2+5\right)+8
   a b^3 v \left(5
   v^2+3\right)-b^4 \left(5
   v^4+10 v^2+1\right)
\big]
  $
  \\ \hline
$(3,2)$
&
  $
  4 \big[
  -2 a^7 v \left(7 v^4+30
   v^2+11\right)
   +5 a^6 b \left(3
   v^6+55 v^4+65 v^2+5\right)
  $\newline$
  -6
   a^5 b^2 v \left(51 v^4+190
   v^2+63\right)+5 a^4 b^3
   \left(17 v^6+265 v^4+275
   v^2+19\right)
$\newline$     
   -10 a^3 b^4 v
   \left(53 v^4+170
   v^2+49\right) 
   +3 a^2 b^5
   \left(19 v^6+255 v^4+225
   v^2+13\right)
$\newline$    
   -10 a b^6 v
   \left(11 v^4+30
   v^2+7\right)+b^7 \left(3
   v^6+35 v^4+25 v^2+1\right)
  \big]
  $
  \\\hline\hline
  $(4,1)$ &
  $
  (105 \pi/16)(a v-b) (a-b v)^3 \big[
  a^4 \left(-8 v^4-20
   v^2-5\right)+12 a^3 b v
   \left(6 v^2+5\right)
  $\newline$
  -2 a^2
   b^2 \left(10 v^4+79
   v^2+10\right)+12 a b^3 v
   \left(5 v^2+6\right)-b^4
   \left(5 v^4+20 v^2+8\right)\big]
  $
  \\ \hline
$(4,2)$
&
  $
  (15 \pi/32)(a-bv)\big[
  -2 a^8 v \left(24 v^6+320
   v^4+485 v^2+95\right)
  $\newline$  
  +7 a^7 b
   \left(248 v^6+1100 v^4+635
   v^2+30\right)-a^6 b^2 v
   \left(760 v^6+15808 v^4+25345
   v^2+4980\right)
  $\newline$
  +105 a^5 b^3
   \left(92 v^6+466 v^4+276
   v^2+13\right)-15 a^4 b^4 v
   \left(106 v^6+2272 v^4+3860
   v^2+769\right)
  $\newline$  
  +21 a^3 b^5
   \left(405 v^6+2050 v^4+1258
   v^2+60\right)-a^2 b^6 v
   \left(585 v^6+12140 v^4+20270
   v^2+4196\right)
  $\newline$
  +7 a b^7
   \left(160 v^6+775 v^4+460
   v^2+24\right)-5 b^8 v \left(4
   v^6+79 v^4+124 v^2+24\right)
  \big]
  $
  \\\hline\hline

$(5,1)$
 & 
$
4(a v-b) (a-b v)\big[
a^8 \left(v^8-36 v^6-378 v^4-420
   v^2-63\right)
$\newline$
+64 a^7 b v
   \left(v^6+27 v^4+63
   v^2+21\right)
   -4 a^6 b^2
   \left(9 v^8+668 v^6+3222
   v^4+2268 v^2+105\right) 
$\newline$
+64
   a^5 b^3 v \left(27 v^6+289
   v^4+405 v^2+63\right)
   -2 a^4
   b^4 \left(189 v^8+6444
   v^6+18094 v^4+6444
   v^2+189\right)
$\newline$    
+64 a^3 b^5 v
   \left(63 v^6+405 v^4+289
   v^2+27\right)
   -4 a^2 b^6
   \left(105 v^8+2268 v^6+3222
   v^4+668 v^2+9\right)    
$\newline$
+64 a b^7
   v \left(21 v^6+63 v^4+27
   v^2+1\right)
   -b^8 \left(63
   v^8+420 v^6+378 v^4+36
   v^2-1\right)
    \big]
    $
\\ \hline

$(5,2)$
 & 
$
    -2 \big[
2 a^{11} v \left(11 v^8+500
   v^6+2114 v^4+1652
   v^2+203\right)    
$\newline$
-7 a^{10} b
   \left(3 v^{10}+467 v^8+4214
   v^6+6734 v^4+2087
   v^2+63\right)
$\newline$
+2 a^9 b^2 v
   \left(1665 v^8+35036
   v^6+110726 v^4+73052
   v^2+8001\right)
$\newline$
-21 a^8 b^3
   \left(51 v^{10}+3491
   v^8+22358 v^6+28910 v^4+7703
   v^2+207\right)
$\newline$
+12 a^7 b^4 v
   \left(2845 v^8+41836
   v^6+103726 v^4+56812
   v^2+5341\right)
$\newline$
-42 a^6 b^5
   \left(133 v^{10}+6517
   v^8+32410 v^6+33922 v^4+7489
   v^2+169\right)
$\newline$
+84 a^5 b^6 v
   \left(829 v^8+9580 v^6+19054
   v^4+8428 v^2+637\right)
$\newline$
-6 a^4
   b^7 \left(1029 v^{10}+40789
   v^8+164122 v^6+137410
   v^4+23617 v^2+393\right)
$\newline$
+42
   a^3 b^8 v \left(795 v^8+7604
   v^6+12194 v^4+4148
   v^2+219\right)
$\newline$
-a^2 b^9
   \left(1449 v^{10}+49049
   v^8+162722 v^6+105770
   v^4+12437 v^2+93\right)
$\newline$
+14 a
   b^{10} v \left(203 v^8+1652
   v^6+2114 v^4+500
   v^2+11\right)
$\newline$
-b^{11} \left(35
   v^{10}+1043 v^8+2870 v^6+1358
   v^4+71 v^2-1\right)
\big]
    $
\\\hline\hline
  $(6,1)$ &
  $
  (3465 \pi/128)(a v-b) (a-b v)^5 \big[
  a^6 \left(-32 v^6-112 v^4-70 v^2-7\right)
  $\newline$
  +26 a^5
   b v \left(16 v^4+28 v^2+7\right)-a^4 b^2
   \left(112 v^6+1796 v^4+1337 v^2+70\right)
   $\newline$
   +52
   a^3 b^3 v \left(14 v^4+57 v^2+14\right)-a^2
   b^4 \left(70 v^6+1337 v^4+1796
   v^2+112\right)
   $\newline$
   +26 a b^5 v \left(7 v^4+28
   v^2+16\right)-b^6 \left(7 v^6+70 v^4+112
   v^2+32\right)
   \big]
  $
  \\ \hline
\end{tabular}

\caption{Scattering angle of a spinning probe in a Kerr background up to 5th order in $G$. \\$\chi_{n,k}$ is the $G^nS^k$ contribution to the full scattering angle $\chi=\sum_{n,k}\chi_{n,k}$.}
\end{table}

\section{Adding Spin to the Probe}

It is well known that it is possible to consider a probe limit in which the mass is negligible but the (rescaled) spin of the probe is finite. In this section we extend the scattering
angle calculation to the case of a spinning probe in a Kerr metric. We will be able to derive results up to and including second order in the probe spin.

The description of the motion of extended bodies in general relativity is a complicated problem, and one usually needs to resort to the use of approximation schemes. For example, one may utilize a multipolar approximation method originally devised by Tulczyjew \cite{Tulczyjew:1959}, to explicitly work out the equations of motion.
This method was applied by Steinhoff and Puetzfeld in ref.~\cite{Steinhoff:2009tk}, using a multipole approximation up to the quadrupolar order, 
i.e. keeping the
covariant monopole $t^{\mu\nu}$, dipole $t^{\mu\nu\alpha}$ and quadrupole 
$t^{\mu\nu\alpha\beta}$ moments. 
They write the stress-energy tensor $T^{\mu\nu}$ in the manifestly covariant form
\begin{equation}\label{Texp}
\sqrt{-g} T^{\mu\nu} = \int \dd \tau \bigg[
	t^{\mu\nu} \delta_{(4)}
	-\nabla_{\alpha} ( t^{\mu\nu\alpha} \delta_{(4)} )
	+ \frac{1}{2!} \nabla_{\alpha\beta}( t^{\mu\nu\alpha\beta} \delta_{(4)} )
\bigg] \,.
\end{equation}
Here $\tau$ is the proper time of the worldline $x^{\rho}(\tau)$, and $\delta_{(4)} \equiv \delta(y^{\rho} - x^{\rho}(\tau))$. The dynamics of the multipolar test body follow from demanding that the stress-energy tensor (\ref{Texp}) is covariantly conserved
\begin{equation}\label{Tvar}
\nabla_{\nu}{T^{\mu\nu}}= 0 \,.
\end{equation}
This is sometimes referred to as Mathisson's variational equations of mechanics \cite{Mathisson:1937,Mathisson:2010} and imposes certain conditions on the multipole moments. Building on Tulczyjew's method \cite{Tulczyjew:1959}, eq.~(\ref{Tvar}) is explicitly evaluated in ref.~\cite{Steinhoff:2009tk} where it was found that the multipole tensors
$t^{\mu\nu\dots}$ can be expressed in terms of a vector $p^{\mu}$, an antisymmetric tensor $S^{\mu\nu}$, and Dixon's reduced moment $J^{\mu\nu\alpha\beta}$, which has the same symmetries as the Riemann tensor. 
In terms of them, the stress-energy tensor becomes \cite{Steinhoff:2009tk}
\begin{equation}\label{Qset}
\sqrt{-g} T^{\mu\nu} = \! \int \! \dd \tau \bigg[
	\dot x^{(\mu} p^{\nu)} \delta_{(4)}
	+ \frac{1}{3} \Riem_{\alpha\beta\rho}{}^{(\mu} J^{\nu)\rho\beta\alpha} \delta_{(4)}
	+ \nabla_{\alpha}\left( \dot x^{(\mu} S^{\nu)\alpha} \delta_{(4)} \right)
	- \frac{2}{3} \nabla_\alpha\nabla_\beta\left( J^{\mu\alpha\beta\nu} \delta_{(4)} \right) \!
\bigg] \!,
\end{equation}
where $\dot x^\mu$ is the tangent to the worldline, $R_{\mu\nu\rho\sigma}$ is the Riemann tensor (defined via $2\nabla_{[\mu}\nabla_{\nu]}w_\rho=R_{\mu\nu\rho}{}^\sigma w_\sigma$), where the (square) brackets denote (anti-) symmetrization of enclosed indices (e.g.\ $A_{(\mu\nu)}=\frac{1}{2}(A_{\mu\nu}+A_{\nu\mu})$). 
The vector $p^{\mu}$ and tensor $S^{\mu\nu}$ are then identified as the linear momentum vector and spin tensor of the object (which now play the role
of monopole and dipole moment). The motion of a multipolar test body (or probe) in a generic curved background spacetime is described by the two equations governing the evolution of its momentum and spin along the worldline, which are also obtained by evaluating eq.~(\ref{Tvar}) in ref.~\cite{Steinhoff:2009tk}.  Through the quadrupolar order in the multipole expansion, they read
\begin{subequations}\label{MPD}
\begin{alignat}{3}\label{Dixonp}
\frac{Dp_\mu}{d\tau}\,+\,\frac{1}{2}R_{\mu \nu \rho\sigma}\dot x^\nu  S^{\rho\sigma}
&=-\frac{1}{6}\nabla_\mu R_{\kappa\lambda\rho\sigma}J^{\kappa\lambda\rho\sigma},
\\\label{DixonS}
\frac{DS^{\mu \nu }}{d\tau}-2p^{[\mu }\dot x^{\nu ]}
&=\frac{4}{3}R^{[\mu}{}_{\lambda\rho\sigma}J^{\nu]\lambda\rho\sigma}.
\end{alignat}
\end{subequations}
These are the Mathisson-Papapetrou-Dixon  equations \cite{Mathisson:1937,Mathisson:2010,Dixon:1974}. They may also be derived from (\ref{Tvar}) without the assumption of a distributional $T^{\mu\nu}$ (see \emph{e.g.}~ref.~\cite{Harte:2014wya}), or alternatively from an effective action (see refs.~\cite{Marsat:2014xea,Vines:2016unv} for recent examples).
For a specific quadrupole tensor given as a function of $p^\mu$ and $S^{\mu\nu}$, a closed set of evolution equations is completed by the imposition of a ``spin supplementary condition''. 
We will here employ the  Tulczyjew-Dixon choice \cite{Tulczyjew:1959,Dixon:1974},
\be\label{SSC}
p_\mu S^{\mu\nu}=0,
\ee
which, together with (\ref{MPD}), determines the worldline tangent $\dot x^\mu$ in terms of the other quantities.  Given a Killing 
vector field $\xi^\mu$ of the background space-time, and regardless of the choice of the spin supplementary condition, an important property of the equations (\ref{MPD}) is that the quantity 
\be\label{Pxi}
\mathcal P_\xi=\xi^\mu p_\mu+\frac{1}{2}S^{\mu\nu}\nabla_\mu\xi_\nu
\ee
is conserved along the worldline, $i.e.$, $D\mathcal P_\xi/d\tau=0$. This holds to all orders in the multipole expansion \cite{Harte:2014wya}.  The system of equations (\ref{MPD})--(\ref{SSC}) is explicitly invariant under reparametrizations of the worldline, but for simplicity we will here adopt the condition $\dot x^2\equiv\dot x_\mu\dot x^\mu=-1$, making $\tau$ the proper time.

A form of the quadrupole tensor $J$ appropriate to describe a spin-induced quadrupole, quadratic in the spin, and assuming eq.~(\ref{SSC}) is given by 
\be\label{DixonJ}
J^{\mu\nu\rho\sigma}=\frac{-3}{(-p^2)^{3/2}}p^{[\mu}S^{\nu]}p^{[\rho}S^{\sigma]},
\ee
for the case of a black-hole probe \cite{Steinhoff:2011sya}. We will restrict ourselves to such probes here but stress that probes with internal and finite-size 
structure can be treated in this formalism as well.  Here, $S^\mu$ is the Pauli-Lubanski spin vector,
\be\label{PauliLub}
S^\mu=-\frac{1}{2}\epsilon^\mu{}_{\nu\rho\sigma}\frac{p^\nu}{\sqrt{-p^2}}S^{\rho\sigma}
\qquad\Leftrightarrow\qquad
S^{\mu\nu}=\epsilon^{\mu\nu}{}_{\rho\sigma}\frac{p^\rho}{\sqrt{-p^2}}S^\sigma,
\ee
with $p_\mu S^\mu=0$. It has invariant magnitude
\be
S^2 \equiv S_\mu S^\mu=\frac{1}{2}S_{\mu\nu}S^{\mu\nu}.
\ee

We next solve for the worldline tangent $\dot{x}^{\mu}$ by covariantly differentiating (\ref{SSC}) with respect to $\tau$ and inserting equations (\ref{MPD}).  For our case of a black-hole probe with its associated spin-induced quadrupole (\ref{DixonJ}), and working perturbatively in the test body's spin $S$, one finds after a remarkable cancellation the simple relation
\be\label{tangentp}
\dot x^\mu=\frac{p^\mu}{\sqrt{-p^2}}+\mathcal O(S^3),
\ee
as noted in ref.~\cite{Bini:2015zya}. That is, the tangent is still proportional to the momentum through this order, for a black hole.  Finally, one can verify from (\ref{MPD}) with (\ref{DixonJ})--(\ref{PauliLub}) that the quantity 
\be\label{msquared}
m^2 \equiv p^2+R_{\mu\nu\rho\sigma}\frac{p^\mu p^\rho}{-p^2}S^\nu S^\rho+\mathcal O(S^3)
\ee
is conserved to the order shown. Taking the flat space limit, we identify $m$ with the mass of the scattered probe.  
All of this holds in a general curved background.

We now restrict ourselves to the background of a Kerr spacetime outside a black hole of mass $M$ and spin $Ma$ 
in Boyer-Lindquist coordinates $x^\mu=(t,r,\theta,\phi)$.  We again consider the motion in the equatorial plane $\theta=\pi/2$, and with the probe spin aligned (or anti-aligned) with the symmetry axis, $S^\mu=-S e_\theta{}^\mu$ where $e_\theta{}^\mu$ is the unit vector in the $\theta$ direction. We take the constant scalar $S$ to carry a sign: positive when the probe spin is aligned with the Kerr spin, and negative when anti-aligned.  Note that the motion will remain in the equatorial plane only when the spin is aligned, and the spin will remain aligned only when the motion is in the equatorial plane.  In this case, the evolution equation (\ref{DixonS}) for $S^{\mu\nu}$ is automatically satisfied, and the content of evolution equation (\ref{Dixonp}) for $p^\mu$ is equivalent to the conservation equations for three constants of motion: the invariant mass $m$ of (\ref{msquared}) and the two constants (\ref{Pxi}) from the two Killing vectors of the Kerr background.  The timelike Killing vector $t^\mu \equiv (\partial_t)^\mu$ gives the conserved energy $E \equiv \mathcal P_t$, and the axial Killing vector $\phi^\mu \equiv (\partial_\phi)^\mu$ gives in this aligned-spin/equatorial case the total angular momentum $J \equiv \mathcal P_\phi$,
\begin{alignat}{3}\label{Eexact}
E&=-p_at^a-\frac{1}{2}S^{ab}\nabla_at_b 
&
\qquad\quad J&=p_a\phi^a+\frac{1}{2}S^{ab}\nabla_a\phi_b
\\\label{Jexact}
&=-p_t+\frac{GMS}{r^3\sqrt{-p^2}}(p_\phi+ap_t),
&
&=p_\phi+\frac{S}{\sqrt{-p^2}}\Big[{-}p_t+\frac{GMa}{r^3}(p_\phi+ap_t)\Big],
\end{alignat}
where the second line has evaluated in terms of the momentum components $p_\mu=(p_t,p_r,p_\theta=0,p_\phi)$ in Boyer-Lindquist coordinates with $\theta=0, p_\theta=0$ and with the spin tensor as specified above.  Similarly evaluating (\ref{msquared}) yields
\be
m^2=-p^2+\frac{GMS^2}{r^3}\bigg[1+3\frac{(p_\phi+ap_t)^2}{r^2(-p^2)}\bigg]+\mathcal O(S^3),
\ee
with
\be\label{pmusquared}
-p^2=-g^{\mu\nu}p_\mu p_\nu=\frac{[(r^2+a^2)p_t+ap_\phi]^2}{r^2\Delta}-\frac{(p_\phi+ap_t)^2}{r^2}-\frac{\Delta}{r^2}p_r^2,
\ee
and $\Delta \equiv r^2+a^2-2GMr$.  Now the system (\ref{Eexact})--(\ref{pmusquared}) can be solved, working perturbatively in $S$, for the momentum components $p_t$, $p_r$ and $p_\phi$ as functions of only the Boyer-Lindquist radial coordinate $r$ and the constants $M$, $a$, $E$, $J$ and $m$.

From the relation (\ref{tangentp}) for the tangent vector, with $p^\mu=g^{\mu\nu}p_\nu$, evaluating the $r$ and $\phi$ components yields 
\begin{alignat}{3}
\dot\phi&=\frac{1}{\Delta\sqrt{-p^2}}\Big[p_\phi-\frac{2GM}{r}(p_\phi+ap_t)\Big]+\mathcal O(S^3),\phantom{\Bigg|}
\\
\dot r&=\frac{\Delta\,p_r}{r^2\sqrt{-p^2}}+\mathcal O(S^3),
\end{alignat}
which can each be expressed as functions of $r$ and the constants of motion from the results above.  Then the scattering angle 
$\chi$ can be computed from
\be
\chi=2\int_{r_m}^\infty d r\,\frac{\dot\phi}{\dot r} - \pi\, .
\ee
From this expression we can immediately make contact with our general formula (\ref{eq:chiintdr}). 
Note that $d\phi/dr$ up to and including $\mathcal O(S^2)$ has the correct form to readily identify $h(r)$. First, 
\begin{equation}
  \frac{d\phi}{dr}=
  \frac{r (l r-2 G \kappa  M)}{p_{r} \left(a^2+r(r-2 G M)\right)^2}
  -\frac{a G \kappa  M S }{mr p_{r} \left(a^2+r (r-2 G M)\right)^2}
  +\frac{G \kappa  M S ^2 (r-2 G M)}{mr^2 p_{r} \left(a^2+r(r-2 G M)\right)^2}+\mathcal O(S^3)
\end{equation}
where for simplicity 
we have introduced $\gamma=E/m$ and $l \equiv L/m$, where $L\equiv J-\gamma S$ is the orbital angular momentum (cf.~eq.~(\ref{Jexact}) as $r\to\infty$), and $\kappa \equiv l-\gamma a$. The radial momentum $p_r$ is the positive root
of eq.~(\ref{pmusquared}),
\begin{equation}
  p_r=\sqrt{\frac{r^2}{\Delta}\left[\frac{[(r^2+a^2)p_t+ap_\phi]^2}{r^2\Delta}
  -\frac{(p_\phi+ap_t)^2}{r^2}+p^2\right]}\, .
\end{equation}
This identifies
\begin{equation}\label{eq:hr_kerrspin}
  h(r)=
  -\frac{r (l r-2 G \kappa  M)}{\left(a^2+r (r-2 G M)\right)^2}
  +\frac{a G \kappa  M S }{mr \left(a^2+r (r-2 G M)\right)^2}
  -\frac{G \kappa  M S ^2 (r-2 G M)}{m^2r^2 \left(a^2+r (r-2 G M)\right)^2}
  +\mathcal O(S^3) ~.
\end{equation}
We can readily use the expression above in normal coordinate systems by means of the transformation (\ref{Kerr-normal}). 
Furthermore, $h(r)$ obeys our requirements of being real analytic on the interval $r\in [r_m,\infty[$ and with
falls-off as $\lim_{r\rightarrow\infty}h(r)\sim 1/r^{n}$, with $n\geq 2$. Therefore, eq.~\eqref{eq_general} can be used. Note that
to each order in $S$, the radial momentum $p_r$ must be expanded correspondingly.  An equivalent form of this integrand was first derived (by the same methods) in ref.~\cite{Bini:2017pee}.

Results up to sixth order in $G$ and up to second order in probe spin $S$ are given in Table III. We note that the pattern of resummation in the Kerr black hole spin is generalized to an overall prefactor of
\begin{equation}
c_{n,k}\equiv\frac{G^nM^n(S/m)^k}{v^{2n}(b^2 - a^2)^{(3n+2k-1)/2}}
\label{spinPF}
\end{equation}
to first ($k=1$) and second ($k=2$) order in the probe spin. It is tempting to conjecture that this pattern will hold to higher
orders ($k >2$) in the probe spin. Furthermore, we observe that the remainders after factorising eq.~(\ref{spinPF}) again shows remarkable structures to linear order in the spin of the probe $S$, i.e. for $k=1$
\begin{align}
\chi_{n,1}/c_{n,1}&=(av-b)(a-bv)\sum_{\ell=0}^{2n-2}a^\ell b^{2n-2-\ell}f_{n,1,\ell}(v)
\qquad \textrm{for odd $n$}, 
\\
\chi_{n,1}/c_{n,1}&=(av-b)(a-bv)^{n-1}\sum_{\ell=0}^{n}a^\ell b^{n-\ell}f_{n,1,\ell}(v) 
\qquad\  \textrm{for even $n$},
\end{align}
where $f_{n,1,\ell}(v)$ are polynomials in $v$ of order $n$ for even $n$ and order $2n-2$ for odd $n$. We have not found any discernible structure for the results at quadratic order in $S$.

\section{Conclusion}

We have derived a simple formula for the scattering angle of massless probes in external black hole metrics. Building on the compact formula presented
in ref.~\cite{Bjerrum-Bohr:2019kec}, we have found a scattering angle expression that straightforwardly handles metrics in any choice of coordinates belonging to a
class we have denoted as normal. In such coordinates the metric enjoys the property of reducing to flat Minkowski metric in polar coordinates when one takes the
limit $G \to 0$. To illustrate, we have derived the scattering angles of massive and massless probes in the metric of a Schwarzschild black hole in Schwarzschild
coordinates. The final scattering angle formula is manifestly free of any dependence on the turning point $r_m$ of the orbit without any need of regularization or
prescription. 

While of interest in itself, the existence of such a compact formula for the scattering angle becomes more important in the case of scattering in the equatorial plane
of Kerr black hole metrics. Choosing standard Boyer-Lindquist coordinates, one notices that the Kerr metric is not in normal form in those coordinates. We show that
a simple transformation of the radial coordinate brings the Kerr metric to normal form and we are then able to rather effortlessly calculate the scattering angle in
this Kerr metric to any desired order in $G$. Interestingly, we find that the resulting expressions all re-sum the dependence on the black hole spin $a$  to all
orders, for any fixed order in $G$. Finally, we have extended these scattering angle calculations to the case of spinning black-hole probes in the aligned (or anti-aligned) 
case of spins in the equatorial plane of the Kerr metric. Our results display regularities up to second order in the
probe spin that may lead to a better understanding of all-order results in the case of scattering of spinning black holes. 
We expect the resulting expressions to be useful for the community presently computing scattering angles from gravitational 
scattering amplitudes in the Post-Minkowskian expansion.

\vspace{1cm}


\begin{acknowledgements}
{\sc Acknowledgements:} 
The work of P.H.D. and A.L. was supported in part by Independent Research Fund Denmark, grant number 0135-00089B. This research was also supported in part by the National Science Foundation under Grant No. NSF PHY-1748958. AL is supported partially by the European Union's Horizon 2020 research and innovation programme under the Marie Sk\l{}odowska-Curie grant agreement  No. 847523 INTERACTIONS. We thank the Kavli Institute of theoretical Physics for hospitality during the program "High-Precision Gravitational Waves" where part of this work was done.

\end{acknowledgements}

\end{document}